\documentclass[aps,prb,twocolumn,showpacs]{revtex4-1} 
\newcommand{\cmo}{\mbox{CaMn$_{7}$O$_{12}$}}

\newcommand{\tna}{\mbox{$T_\mathrm{N1}$}}
\newcommand{\tnb}{\mbox{$T_\mathrm{N2}$}}
\usepackage[normalem]{ulem}
\usepackage{amsmath}
\usepackage{graphicx}
\usepackage{xcolor}
\usepackage{bm}% bold math
\begin{document}

\title{Magnetic and orbital correlations in multiferroic CaMn$_{7}$O$_{12}$ probed by x-ray resonant elastic scattering}
\author{K. Gautam$^{1}$}\author{S. S. Majid$^{2}$}\author{S. Francoual$^{3}$}\author{A. Ahad$^{2}$}\author{K. Dey$^{1}$}
\author{M. C. Rahn$^{4}$}\author {R. Sankar$^{5,6}$}\author {F.C. Chou$^{6}$}\author{D. K. Shukla$^{1}$}
\email{dkshukla@csr.res.in}

\affiliation{ $^{1}$UGC-DAE Consortium for Scientific Research, Khandwa Road, Indore 452001, India}

\affiliation{ $^{2}$Department of Physics, Aligarh Muslim University, Aligarh 202002, India}

\affiliation{$^{3}$Deutsches Elektronen-Synchrotron, Notkestrasse 85, D-22607 Hamburg, Germany}

\affiliation{$^{4}$Department of Physics, Clarendon Laboratory, University of Oxford, Oxford OX1 3PU, United Kingdom}
\thanks{Current address: Institute for Solid State and Materials Physics, Technical University of Dresden, 01062 Dresden, Germany}

\affiliation{$^{5}$ Institute of Physics, Academia Sinica, Taipei, 11529, Taiwan}

\affiliation{$^{6}$Center for Condensed Matter Sciences, National Taiwan University, Taipei 10617, Taiwan}
                   
%\affiliation{$^{6}$ Advanced Photon Source, Argonne National Laboratory, Argonne, Illinois 60439, USA}

\date{\today}
    
\begin{abstract}
The quadruple perovskite $\cmo$\ is a topical multiferroic, in which the hierarchy of electronic correlations driving structural distortions, modulated magnetism, and orbital order is not well known and may vary with temperature. x-ray resonant elastic scattering (XRES) provides a momentum-resolved tool to study these phenomena, even in very small single crystals, with valuable information encoded in its polarization- and energy-dependence. We present an application of this technique to \cmo. By polarization analysis, it is possible to distinguish superstructure reflections associated with magnetic order and orbital order. Given the high momentum resolution, we resolve a previously unknown splitting of an orbital order superstructure peak, associated with a distinct \textit{locked-in} phase at low temperatures. A second set of orbital order superstructure peaks can then be interpreted as a second-harmonic orbital signal. Surprisingly, the intensities of the first- and second-harmonic orbital signal show disparate temperature and polarization dependence. This orbital re-ordering may be driven by an exchange mechanism, that becomes dominant over the Jahn-Teller instability at low temperature.

\end{abstract}


\maketitle 

%\section{Introduction}
\label{1}

In materials with strong electronic correlations, the coupling of spin, orbital and lattice degrees of freedom can give rise to unusual phenomena, like colossal magneto-resistance, chiral orders, metal-insulator transitions, superconductivity and multiferroicity. While the resulting bulk physical properties are often not straightforward to interpret, in many cases a microscopic (i.e., momentum- and energy-resolved) understanding was achieved through x-ray resonant elastic scattering\cite{wilkins2003soft,paolasini2002coupling,blanco2013momentum,blake2001transition}. In multiferroics, the coexistence and coupling of several order parameters can obscure the roles of the different electronic interactions that are at play, but also creates unusual scenarios in which novel emergent phases may arise. Notably, ordered phases in such materials often couple strongly to external fields, which adds potential for technological applications, for example, as in magneto-electric random access memory and multiferroic tunnel junctions~\cite{scott2007data,pantel2012reversible}.

In $\cmo$, a cubic ($Im\bar{3}$) to rhombohedral ($R\bar{3}$) structural phase transition occurs below 480 K and incommensurate orbital order (OO, propagation vector $\mathbf{q}_{oo}$) develops at $T_{oo}\sim 250$\,K. At $\tna\sim90$\,K, this is followed by the onset of modulated magnetic order (propagation vector $\mathbf{q}_{m1}$), coincident with an electric polarization of the material. In fact, among type-II multiferroic materials, $\cmo$ develops the largest known polarization induced by magnetic order~\cite{scott2007data}. Down to $\tnb\sim 48$\,K, the periodicity of the incommensurate magnetic order is exactly twice that of the orbital modulation ($\mathbf{q}_{m1} = \mathbf{q}_{oo}/2 $), signifying an intimate coupling between the arrangements of orbitals and magnetic moments~\cite{NJ2012,johnson2016modulated,S2008}.

The orbital ordering phenomenon in $\cmo$ is unusually complex due to presence of the Jahn-Teller (JT) active Mn$^{3+}$~($d^4$) ions on two different Wyckoff sites. Each of these sublattices develop a distinct type of distortion: (i) the $A$-type Mn$^{3+}$ ions (Wyckoff sites $9e$ in $R\bar{3}$ are situated at the center of a rhombic prism resembling an elongated octahedron, whereas (ii) the $B$-type Mn$^{3+}$ ions (Wyckoff site $9d$) are contained in an apically contracted octahedral environment -- a rare case among JT distortions~\cite{bochu1980bond}. This combination of magnetic environments with different symmetries makes it difficult to recognize the causality between the resulting orbital, spin, and structural instabilities.

As in other manganites~\cite{tokura2000orbital} there are two candidate mechanisms for OO in $\cmo$: (i) the JT distortion, in which a spontaneous symmetry breaking energetically lowers the occupied crystal-field--split $d$-electronic states, and (ii) an exchange mechanism, in which the crystal relaxes to accommodate couplings between magnetic spin and orbital pseudo-spin degrees of freedom (the pseudo-spin $\tau=\pm1/2$, corresponding to the choice between occupying $d_{x^2-y^2}$ or $d_{z^2-r^2}$ orbitals). Without a detailed knowledge of the material, it is hardly possible to distinguish these mechanisms by bulk measurements, especially if \textit{both} are active in the material~\cite{wilkins2003direct,elfimov1999orbital,benfatto1999critical}. Its capability to disentangle magnetic, orbital, and charge-scattering contributions make XRES a particularly powerful tool to identify which is the dominant effect~\cite{benfatto1999critical,di2003resonant}.

\begin {figure}[htb]
\centering
\includegraphics[width =1\columnwidth]{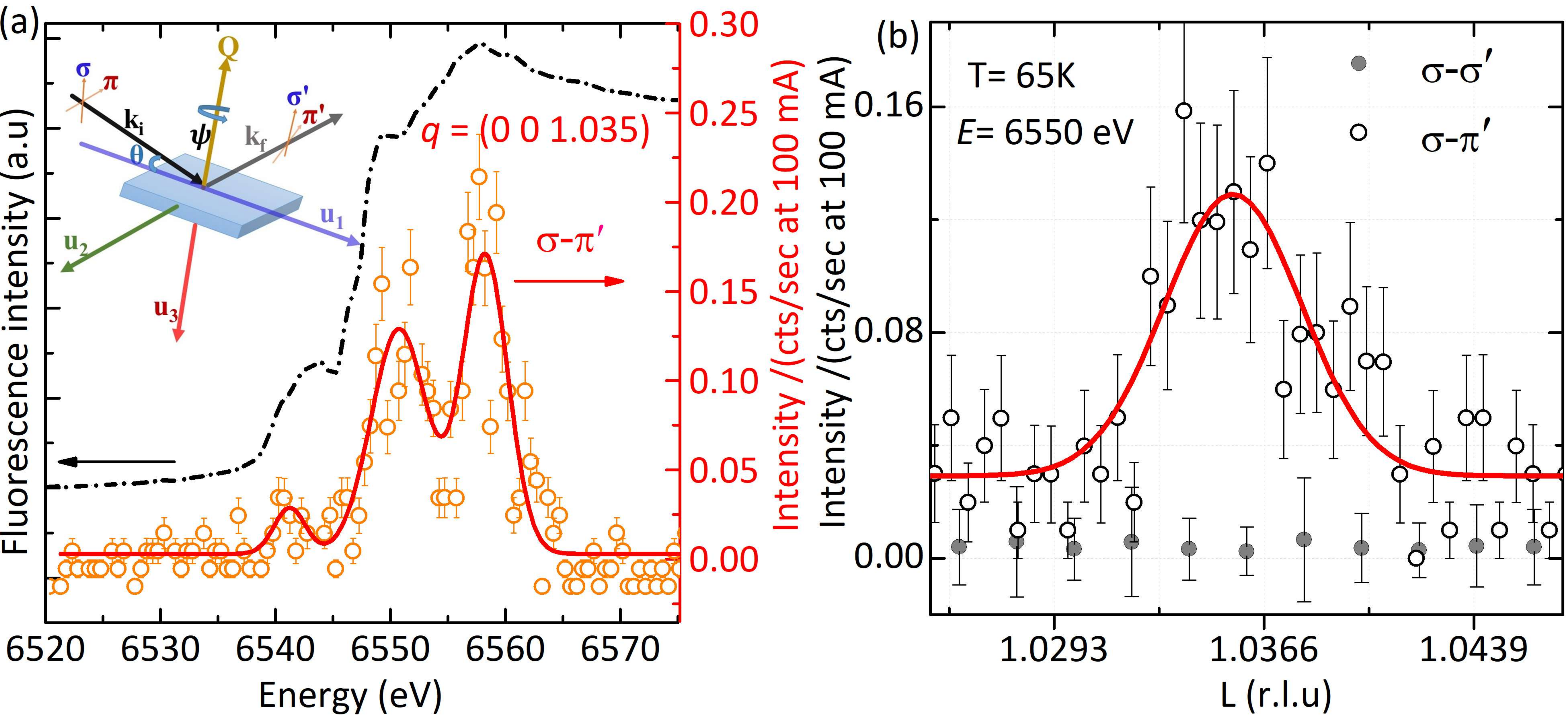}
\caption{\label{Fig1a} (a) x-ray fluorescence across the Mn K edge in $\cmo$, and a corresponding energy scan of the magnetic $\mathbf{q}_{m1}=(0,0,1.035)$ reflection at 65\,K in the $\sigma\pi'$ polarization channel. (b) $L$-scans across this reflection in the $\sigma\pi'$ and the $\sigma\sigma'$ channels at 65\,K and 6550\,eV, confirming its magnetic origin.}
\label{mag}%
\end{figure}
In $\cmo$, the thermal variation of Mn$^{3+}$--O bond lengths can serve as a measure for the relevance of the JT mechanism \cite{przenioslo2004charge}. Between room temperature and $\tna$, they are indeed known to show relatively strong variations. On the other hand, below $\tna$, the Mn--O bond lengths remain almost constant. Thus, while both mechanisms are likely active in $\cmo$, JT distortions can only be the driving instability at high temperatures, and onset of magnetism and electric polarization at $\tna$ is associated with a changeover in the character of the orbital instability.

The spin and orbital structures of $\cmo$ have previously been investigated by neutron (NPD) and x-ray powder diffraction\cite{JO2012,slawinski2012,WS2012,S2008}. Although much information is lost in the polycrystalline average, these techniques are each sensitive to two of four (charge, orbital, structural and magnetic) ordering channels. Taken together, these insights therefore already provided valuable insights into the ordering phenomena in this material (see related text and references ~\cite{S2008,W2009,Smaalen2007,slawinski2012,johnson2016modulated,Z2011} therein in the Supplemental Material\cite{Sup}).

In XRES, the photon energy is tuned to the proximity of atomic absorption edges, which enhances magnetic and multipolar scattering cross sections by orders of magnitude. The technique thus provides a powerful means to determine both magnetic structures and OO. Due to the high momentum-resolution, the spectroscopic selectivity of the resonant process and its polarization dependence, XRES is highly suited to disentangle the origins of different scattering contributions, which makes it the ideal tool to study multiferroics~\cite{murakami1998direct,nandi2008nature}. Here, we report an application of Mn K-edge XRES on a minute ($\approx100\,\mu$m) single crystal of $\cmo$. We track in detail the thermal variation of different superstructure peaks and assign their origin by analyzing their energy and polarization dependence.% {\color{red}>must add some \textit{specific} statements on the conclusions of this study. Something like "signature of dominant mechanisms have been identified" is too vague< }. 

High-quality \cmo\ crystals were grown by the self-flux method, as reported in our earlier study~\cite{gautam2017large}. Details on the crystal structure, characterization, and alignment are provided in the Supplemental Material (see Fig. S1 and Ref.~\cite{1978phase} along with related text\cite{Sup}). A $c$-axis--specular \cmo\ platelet was mounted on the cold finger of a CCR cryostat and probed in the vertical scattering geometry, using the six-circle diffractometer at beamline P09, PETRA III (DESY)~\cite{strempfer2013resonant}. The x-ray energy was tuned to the Mn K edge ($\sim6.5$\,keV) and a Cu(110) crystal was used as an analyzer. Compound refractive lenses focused the beam to $\sim20\,\mu$m at the sample position. In the following, $\sigma$ ($\sigma'$) and $\pi$ ($\pi'$) denote incident (scattered) x-ray polarization perpendicular and parallel to the scattering plane, respectively [see inset to Fig.~\ref{Fig1a}(a)]. Magnetization measurements as a function of temperature and magnetic field were performed in a commercial 7\,T SQUID vibrating sample magnetometer (Quantum Design).

In the rhombohedral ($R\bar{3}$) phase of $\cmo$, the reflection condition for $(0,0,L)$ peaks is $L=3\,n$ (integer $n$). Below $\tna$, $\theta$-$2\theta$ scans along the specular $(0,0,L)$ direction reveal a magnetic superstructure peak at $\mathbf{q}_{m1}=(0,0,1.035)$, in agreement with neutron studies~\cite{slawinski2012,johnson2016modulated}. Figure~\ref{Fig1a} shows energy- and polarization-dependent scans that confirm the magnetic origin of this peak (all measurements at 65\,K). Panel (a) shows the energy dependence of the integrated magnetic ($\sigma\pi'$) intensity across the Mn K edge, along with the x-ray fluorescence signal. Three resonances are observed, corresponding to a pre-edge (6539\,eV) $1s\rightarrow3d$ quadrupole transition (indicative of broken local inversion symmetry), and two $1s\rightarrow4p$ dipole transitions (at 6550 and 6559\,eV, corresponding to the two distinct Mn sites). Even though it couples directly to the $3d$ valence shell, the quadrupole resonant enhancement is only around one fifth of that at the dipole transitions. Figure~\ref{Fig1a}(b) illustrates that the $\mathbf{q}_{m1}$ intensity at 6550\,eV is observed entirely in the crossed polarization channel, a characteristic of magnetic XRES. In agreement with the neutron diffraction study by Johnson~\textit{et al.}, the characterisitcs of this reflection show no thermal variation between $\tna$ and $\tnb$~\cite{johnson2016modulated}.

\begin {figure}[htb] %band
\centering
\includegraphics[width = 1\columnwidth]{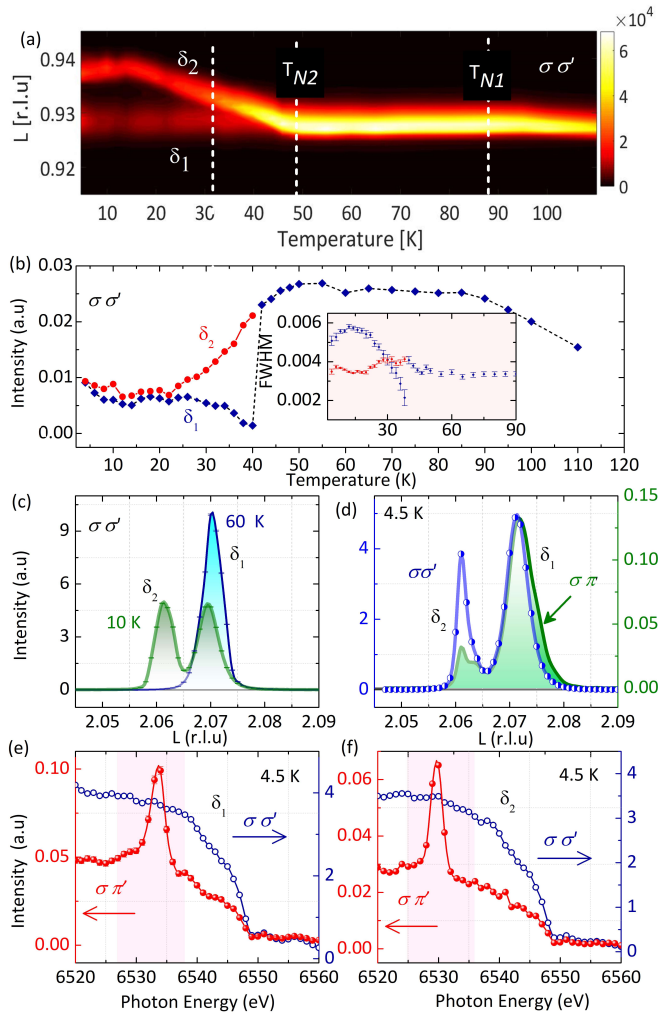}%
\caption{\label{Fig2a} Characteristics of the XRES response from orbital order. (a) Thermal variation of $L$ scans of the $\delta$$_1$ and $\delta$$_2$ orbital superstructure peaks, measured in the $\sigma\sigma'$ channel. (b) Corresponding changes in integrated intensity and linewidths (inset) of these reflections. (c) $L$ scans across the (0,0,2.07) peak in $\sigma\sigma'$ at 10\,K and 60\,K. (d) $L$ scans across this reflection at 4.5\,K, comparing the $\sigma\sigma'$ and $\sigma\pi'$, channels. (e) Energy scans of the (0,0,2.07) intensity, at 4.5\,K, measured in $\sigma\sigma'$ and $\sigma\pi'$. (f) Energy scans at (0,0,2.06), at 4.5\,K, in $\sigma\sigma'$ and $\sigma\pi'$. }
\end{figure}

Apart from the magnetic $\mathbf{q}_{m1}=(0,0,1.035)$ reflection, we also characterized $\mathbf{q}_{oo}=(0,0,L\pm\delta)$  superstructure reflections ($\delta\sim0.93$), which had also been observed earlier in x-ray and neutron studies~\cite{johnson2016modulated,slawinski2012,S2008}. In contrast to $\mathbf{q}_{m1}$-type peaks, these reflections are observed both in the $\sigma\sigma'$ and $\sigma\pi'$ channels. This corroborate that they are not due to a magnetic superstructure, but to a periodic anisotropy in the occupation of the Mn $3d$ $e_g$ ground state manifold. The intensity map in Fig.~\ref{Fig2a}(a) illustrates the thermal variation of $L$ scans around (0,0,0.93) in $\sigma\sigma'$. Evidently, the associated order already exists at temperatures above the onset of magnetism ($\tna=90$\,K), and its modulation length is not affected down to $\tnb=48$\,K. Then, at $T_\mathrm{N2}$, $\mathbf{q}_{s}$ begins to split into a pair of propagation vectors that eventually separated by $\sim0.01$\,reciprocal lattice units. This subtle effect had not been resolved in earlier studies~\cite{johnson2016modulated}.

We label these new branches by their values at low temperatures, $\delta_1=0.93$ ($\approx\delta$), and $\delta_2=0.94$. Their disparate temperature dependence suggests that each may be associated with a different underlying order parameters. While $\delta_1$ does not deviate from the higher-temperature value of $\delta$, $\delta_2$ below $T_\mathrm{N2}$ increases with a constant slope and, at 20\,K, locks into a new constant value. In Fig.~\ref{Fig2a}(b) we illustrate the characteristic evolution of spectral weight and correlation length of the two branches. Below $T_\mathrm{N2}$, in $\sigma\sigma'$ most spectral weight is continuously transferred to the $\delta_2$ reflection. In the low-temperature \textit{decoupled} state, as it is de-locked from the constant $\delta_1$, the $\delta_2$ modulation retains a constant correlation length, while that of $\delta_1$ increases significantly (see inset).

As a check for consistency, we also investigated this splitting on the satelites of another reflection $(0,0,2.07)=(0,0,3-\delta)$ and found the same behavior. In Fig.~\ref{Fig2a}(c), we show $\sigma\sigma'$ $L$ scans across (0,0,2.07), at 10\,K and 60\,K. As at (0,0,0.93), the reflection splits into $\delta_1$ and $\delta_2$ components at low temperatures (here corresponding to $L=2.07$ and $L=2.06$, respectively). We denote this state below $\sim20$\,K as \textit{locked-in state II} (liII), in analogy to the higher-temperature ($\tnb<T<\tna$) locked-in state liI, where $\mathbf{q}_{m1}=\mathbf{q}_{oo}/2$.

The liII regime features a reduction in $\sigma\pi'$ intensity at $\delta_2$, as well as a significant change in the relative intensities between the two peaks ($\delta_1/\delta_2$), in both $\sigma\sigma'$ and $\sigma\pi'$. As shown in Fig.~\ref{Fig2a}(d), at 4.5\,K, the relative intensity is three times higher in $\sigma\pi'$ compared to $\sigma\sigma'$. In Figs.~\ref{Fig2a}(e) and 2(f), energy scans of the $\delta_1$ and $\delta_2$ peaks, in both polarization channels, are shown for the liII phase. A resonance appears in $\sigma\pi'$, which indicates that this is not mere Thompson (charge) scattering, while non-resonant character of the $\sigma\sigma'$ channel implies that Thompson scattering may yet contribute. Similar behavior is observed in the energy scans of $\delta$ in other phases (at $\tna<T$, and $\tnb<T<\tna$)(see Fig.~S3\cite{Sup}). Notably, the resonance energies $E_\mathrm{res}^{\delta1}\sim6534$ and $E_\mathrm{res}^{\delta2}\sim6530$ of these modulations both lie well below the magnetic resonance $E_\mathrm{res}^{m1}\sim6550$\,eV. This is an indication that OO in \cmo\ is driven by a JT distortion which could only move the orbital energy levels ~\cite{castleton2000orbital}.

The resonant behavior discussed above reveals that combined effects of charge (Thompson) and resonant orbital scattering are observed at both $\delta_1$ and $\delta_2$. This supports the previous hypothesis that the stabilization of the $d_{x^2-y^2}$ OO on the Mn$^{+3}$ ($9d$) site (coupled to the oxygen displacements modes of the JT distortion) is driven by charge order on both Mn$^{+3}$ ($9d$) and Mn$^{+4}$ ($3b$) sites~\cite{NJ2012,belik2017reentrant}. Furthermore, a monotonous increase in intensity of the $\delta$ peak, while cooling down to $\tna$ [see Fig.~\ref{Fig2a}(a) and 2(b)], indicates that the contribution of the JT mechanism to this effect reaches its peak at this temperature, and then stays constant down to $\tnb$. This is followed by the bifurcation at $\tnb$, which marks the onset of a new dominant OO mechanism on the background the pre-existing orbital and magnetic order.

\begin {figure}[htb]
\centering
\includegraphics[width = 1\columnwidth]{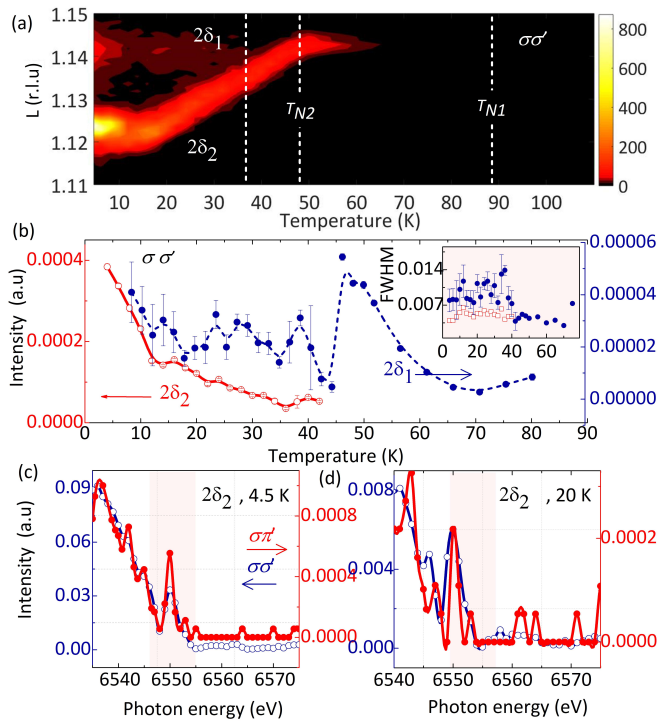}
\caption{(a) Temperature dependence of $L$ scans of the orbital order reflections at $L=3-2\delta_1=1.12$ and $L=3-2\delta_2=1.14$, in the $\sigma\sigma'$ polarization channel. (b) Corresponding variation of integrated intensities and linewidths (see inset). (c) Fixed momentum transfer energy scans of the $L=3-2\delta_1$ peak, measured in both polarization channels, at 4.5 K, and (d) at 20\,K.}
\label{Fig3a}
\end{figure}

In their single crystal neutron diffraction study, Johnson \textit{et al.}~\cite{johnson2016modulated} observed a reflection at $(0,0,k_0)=(0,0,1.12)$ (at 2\,K), and interpreted this as the fundamental magnetic wave vector in the ground state of \cmo. We also confirmed the presence of this peak (at base temperature, $\sim4.5$\,K) by XRES, but find significant intensity not only in $\sigma\pi'$, but also in the $\sigma\sigma'$ channel, which is at odds with purely magnetic order. The similarities of the thermal variations presented here suggest that the peak at $(0,0,k_0)$ actually derives from that at (0,0,1.14): It can be indexed as a second order orbital reflection, $(0,0,3-2\delta$) and first arises below $\tna$ (i.e., in the liI phase). Such higher-harmonic reflections have previously been observed when an incommensurate magnetic structure is accompanied by structural distortions, allowing quadratic magnetoelastic coupling~\cite{strempfer2008magnetic}. As shown in Fig.~\ref{Fig3a}(a), below $\tnb$, the (0,0,1.14) peak splits into $(0,0,3-2\delta_1)=(0,0,1.14)$ and $(0,0,3-2\delta_2)=(0,0,1.12)$, in a similar fashion as at the first-harmonic [cf. Fig.~\ref{Fig2a}(a)].

Figures ~\ref{Fig3a}(a) and ~\ref{Fig3a}(b) illustrate that below $T_\mathrm{N2}$, the new $2\delta_2$ branch gains continuously in $\sigma\sigma'$ spectral weight, while the $2\delta_1$ instability maintains the constant modulus and intensity of the $\delta$ modulation vector. Below $\sim10$\,K, both of these second harmonic reflections appear to be strongly enhanced, indicating higher order (orbital and magnetic) reflection coupling processes in the ground state of this compound. The inset to Fig.~\ref{Fig3a}(b) shows that the correlations lengths of $2\delta_1$ and $2\delta_2$ evolve in parallel to their parent reflections [cf. Fig.~\ref{Fig2a}(b)].

Energy scans of $2\delta_2$, in both polarization channels, are shown in Figs.~\ref{Fig3a}(c) and 3(d), for 4.5\,K and 20\,K, respectively. Against expectations, the resonant behavior of the second-order reflections differs from that at $\delta_1$ and $\delta_2$: The $\sigma\sigma'$ and $\sigma\pi'$ signals are perfectly proportional to each other. As a common feature of these energy scans, the signal is suppressed at the absorption edge, which can be explained in terms of the structural modulation caused due to the displacement of the Mn$^{+4}$ ions (as a consequence of the JT crystal distortion). This was also observed in other manganites~\cite{wilkins2003charge}.

Moreover, the intensities of the first- and second-harmonic OO peaks, also show an opposite thermal variation, adding to the evidence that the first- and second-order intensities are (dominantly) of different origin, even though they are locked into the same underlying periodicity.

In the context of these findings, it appears to be a valid interpretation that the first-harmonic OO peaks are specific to JT-like distortion (they set in at 250\,K $\gg\tna$, with significantly shifted resonance energy), while the second-harmonic signal that appears below $\tna$ is associated with an exchange (magneto-orbital) coupling mechanism. This scenario is also consistent with anomalies in the magnetic entropy that were recently reported by Parul~\textit{et al.}~\cite{jain2018observation} for the temperature regime in which the 2$\delta$ reflection sets in ($\sim 65$\,K).

\begin {figure}[htb]
\centering
\includegraphics[width = 1\columnwidth]{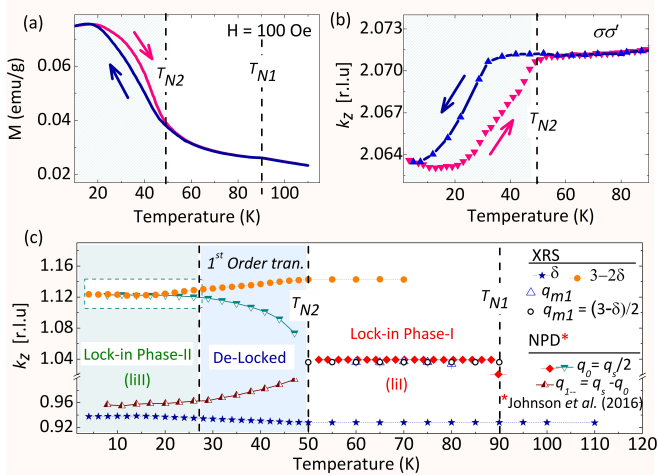}
\caption{(a) Temperature dependence of the magnetization of polycrystalline \cmo\ measured at 100\,Oe, in cooling and warming ramps (see arrows). (b) Temperature dependence of the peak positions of the $L$ scans of the $\sigma\sigma'$ intensity around $k_z\sim2.07$ (arrows indicate cooling and warming ramps). (c) Phase diagram summarizing the temperature dependence of Fourier components of orbital and magnetic order in \cmo. Along with the present (XRES) findings, we reproduce the neutron diffraction results by Johnson~\textit{et al.}~\cite{johnson2016modulated}. Two lock-in phases (liI and liII) can be identified, separated by a broad first order phase transition, during which which the modulation lengths of orbital and magnetic order are de-locked.}%
\label{Fig4a}
\end{figure}

In the following, we discuss the entanglement of order parameters below $\tnb$. Slawinski~\textit{et al.}~\cite{W2009}, using NPD, discovered a weak inflection of the magnetic propagation vector near 37\,K, but were not able to achieve a reliable refinement of the modulus of the ordered magnetic moment between 30\,K and 48\,K. As seen in the magnetization curves shown in Fig.~\ref{Fig4a}(a), we observe a hysteresis between cooling and warming ramps in a similar temperature regime (20\,K -- 48\,K), which can be interpreted as a first order magnetic order-order transition. Figure~\ref{Fig4a}(b) illustrates that the same hysteresis is also manifest in the modulation length $k_z$ of the (0,0,2.07) reflection. Aside from the de-locking of orbital and spin modulations, the appearance of a new orbital reflection (2$\delta$) just above this hysteretical regime underlines the intricate relation to orbital-reordering and indicates that orbitals play a key role in this order-order phase transition.

In a theoretical study of this low temperature regime, Dai \textit{et al.}~\cite{dai2015ferroelectric} concluded that the structural symmetry in the ground state of $\cmo$ should be described by the space group $P3$, rather than $R\bar{3}$. Kadlec~\textit{et al.}~\cite{kadlec2014possible}, observed a dielectric anomaly near~20 K and changes in the far IR transmittance below $\tnb$. Based on the Landau theory of phase transition, Johnson~\textit{et al.}~\cite{johnson2016modulated} also associated the dielectric anomalies below $\tnb$ with a first-order transition. In the case of structural transition from $R\bar{3}$ (centro-symmetric) to $P3$ (noncentro-symmetric) below $\tnb$, the Wyckoff positions should split from $9d$ (B-site Mn$^{3+}$) into two ($3f$ and $6g$) and three ($3d$, $3d$, and $3d$) independent sites. The Mn$^{3+}$ ions at these sites would then be in different orbital states and might give rise to different types of incommensurate structural modulations. The observed splitting of OO reflections below $\tnb$ could be the consequence of such a structural transition. Further dedicated experimental and theoretical studies are required to address this possibility for
understanding the exact origin of these splittings.\\
Taken together, the present results allow us to establish a relation between the orbital and the magnetic modulations at low temperatures. The phase diagram shown in Fig.~\ref{Fig4a}(c) provides an overview of the superstructure Fourier components that have so far been observed in $\cmo$. The propagation vectors determined in the present XRES study are shown, along with the neutron scattering results by Johnson~\textit{et al.} (reproduced from Ref.~\cite{johnson2016modulated}). As discussed above, we identify two locked-in phases (liI and liII) that are separated by a de-locked regime. In the liI phase ($T_\mathrm{N2}<T<T_\mathrm{N1}$), spin and orbital modulations are constant and locked to each other by the relation $\mathbf{q}_{oo} = \mathbf{q}_{m1} = (0,0,3-\delta)/2$ (with $\delta\sim0.93$), which is confirmed by the perfect consistency of the neutron and x-ray results. No such relation seems to exist in the de-locked phase (30\,K $<T<T_\mathrm{N2}$), where $\mathbf{q}_{m1}$ splits into $\mathbf{q}_0$ and $\mathbf{q}_{1-}$~\cite{johnson2016modulated}. In the liII phase, $\mathbf{q}_{1-}$ locks into $(0,0,3-2\delta)$ and thus a fixed relation between the magnetic and orbital modulations is then re-established. %This indicates a re-entrant direct coupling between the spin and orbital order.

 %Previously, Johnson \textit{et al.}\cite{johnson2016modulated} showed a third harmonic magnetic (3k$_o$) wave vector in between $\tna$ and $\tnb$ which is an spin density wave (SDW) with an identical polarization to the primary SDW and indicates a periodic variation in the local spin helicity due to magneto-orbital coupling. In our XRES measurements we have observed second harmonic [(003)-2$\delta$)] OO reflections. These higher harmonic reflections are magnetostrictive in nature and arise from quadratic magneto-elastic coupling between spin and orbital density wave. At temperatures in between $\tnb$ down to $\sim30$ K, first order phase transition region, 

In summary, we have used resonant x-ray elastic scattering to investigate the correlation of magnetic and orbital modulations in multiferroic $\cmo$. It is clear that orbital degrees of freedom play a leading role in the emergence of magnetic order and the large spontaneous electric polarization below $\tna$. We found that the orbital and magnetic modulations are coupled in between $\tna$ and $\tnb$ through the relation $\mathbf{q}_{m1}=\mathbf{q}_{oo}/2=(0,0,3-\delta)/2$, where $\delta = 0.93$. Between 20\,K to $\sim 45$\,K, these modulations are incommensurate with respect to each other, corresponding a very broad orbital-involved order to order transition. Our direct observation of this first order phase transition also provides an explanation for a number of anomalies that had been observed in bulk measurements~\cite{W2009,kadlec2014possible,dai2015ferroelectric}.

At low temperatures in liII, magnetic and orbital modulations couple again, with the previously unknown relation of propagation vectors $(0,0,k_0)=(0,0,3-2\delta)$. Our measurements of the photon energy dependence of the resonant x-ray scattering intensity allows us to disentangle the dominant driving mechanisms of the orbital orderings (purely electronic versus JT distortion driven) near the two multiferroic transition temperatures. Orbital ordering at the wave vector $\delta$ is dominated by the JT mechanism, while the previously unknown higher harmonic $2\delta$ is driven by magneto-orbital exchange interactions. The observation of second-harmonic superstructure reflections proves the existence of a quadratic magnetoelastic coupling below $\tna$ in $\cmo$.\\

K.G. acknowledges support from the Council of Scientific and Industrial Research, New Delhi, India, in the form of SRF. D.K.S. is grateful for support through a DST INSPIRE faculty award (No.IFA-13/PH-52), and to the DST-DESY project for financial support for synchrotron x-ray resonant magnetic scattering experiments. The authors also extend their thanks to R.J. Choudhary for providing magnetization measurements and to G. Panchal for his assistance in one of the synchrotron experiments. R.S. acknowledges financial support provided by the Ministry of Science and Technology in Taiwan under Project No. MOST-108-2112-M-001-049-MY2. RS and FCC both acknowledge Academia Sinica for the budget of AS-iMATE-109-13. Part of this research was carried out at the PETRA III synchrotron radiation source at DESY, a member of the Helmholtz Association. 

\bibliography{cmo_bib}
\end{document}